\documentclass[a4paper,10pt,twoside]{cpc-hepnp}
\usepackage{multicol}
\usepackage{mathtools}
\usepackage{multicol}
\usepackage{graphicx}
\usepackage{booktabs}
\usepackage{caption}
\usepackage{amssymb,bm,mathrsfs,bbm,amscd}
\usepackage{amsmath}
\usepackage{CJK}
\usepackage{times}
\usepackage{color}
\begin{document}
\begin{CJK*}{GBK}{song}

\title{Constraints on $H^\pm$ parameter space in 2HDM at $\sqrt{s}=$ 8 TeV and $\sqrt{s}=$ 13 TeV}

\author{%
Ijaz Ahmed$^{1}$\email{ijaz.ahmed@riphah.edu.pk} %
\quad Murad Badshah$^{1}$\email{muradbadshah25295@gmail.com } %
\quad Nadia kausar$^{1}$\email{Nkausar430@gmail.com} %
}

\maketitle

\address{%
$^{1}$ Riphah International University, Hajj Complex, I-14 Islamabad \\
}
\begin{abstract}
This paper reflects the heavy Higgs scenario where the mass of charged Higgs is equal to or greater than 200 GeV. The CMS observed and expected values of upper limits on the product $\sigma_{H^\pm} \times BR(H^\pm \rightarrow tb^{\pm})$, assuming $H^\pm \rightarrow tb^{\pm}=1$, both at 8 TeV (at integrated luminosity of 19.7 $fb^{-1}$) and 13 TeV (at integrated luminosity of 35.9 $fb^{-1}$ ) center of mass energies are used. By comparing these expected and observed upper limits with computational values, it is found out the expected and observed exclusion regions of charged Higgs parameter space ($ m_H^\pm$ - tan$\beta$ space) in 2HDM both at $\sqrt{s}=$ 8 TeV and $\sqrt{s}=$ 13 TeV. We compare the expected and observed exclusion regions and observe that exclusion regions made by observed upper limits are always greater than the exclusion made by expected upper limits both at 8 TeV and 13 TeV energies. Only in the mass range from 200 GeV to 220 GeV the expected exclusion region is greater than the observed one only at $\sqrt{s}=$ 13 TeV. We also equate the exclusion regions at these two different center of mass energies and find that the expected exclusion region and observed exclusion region at $\sqrt{s}=$13 TeV are always greater than the expected exclusion region and observed exclusion region at $\sqrt{s}=$ 8 TeV respectively.
\end{abstract}

\begin{multicols}{2}

\section{Introduction}
Particle physics deals with the study of elementary / fundamental particles and the interactions which are arising among these particles. Now, the question is, how the problems arising during the interactions of elementary particles are solved? The answer is , Standard Model (SM). SM solves almost all problems of particles at low energies. At high energies the standard model fails as Newtonian mechanics fails at relativist velocities and then we look beyond standard model like MSSM (minimal supersymmetric standard model), 2HDM (two Higgs doublet model) etc.
In SM there is only one Higgs doublet while in 2HDM there are two Higgs doublets. In 2HDM there are total five physical Higgs particles, one is CP even heavy Higgs (H) , one is CP even light Higgs (h) , one is CP odd Higgs (A) and two are charged Higgs ($H^\pm$). 
The masses of $H^\pm$ ($m_{H^\pm}$) are divided into three categories w.r.t the  top (t) quark's mass ($m_t$). If $m_{H^\pm}$ is very small as \textcolor{red}{compared} to $m_t$ (173 GeV) then it is said to be light $H^\pm$  particle , if $m_{H^\pm}$ is comparable to $m_t$ then it is called intermediate $H^\pm$ and if $m_{H^\pm}$ is very large as \textcolor{red}{compared} to $m_t$ then it \textcolor{red}{is} said to be heavy $H^\pm$ particle. Moreover the alignment limit $-\sin{(\alpha-\beta)}$ is taken to be equal to unity through the paper. Where $\alpha$ and  $\beta$ are the mixing angles, as in  this alignment limit the neutral Higgs\textcolor{red}{,} h of 2HDM behaves like standard model Higgs / discovered Higgs. Because if $-\sin{(\alpha-\beta)} \to$ 1 or if $\cos{(\alpha-\beta)} \to$ 0  then h will be similar to SM Higgs but if $\cos{(\alpha-\beta)} \to$ 1 or $-\sin{(\alpha-\beta)} \to $ 0 then H  will be similar to SM Higgs. Discussing these alignment limits are very important because both h and H are CP even and unfortunately the discovered Higgs (SM like) is also calculated to be CP even , so then it becomes difficult to decide which one between h and H is the discovered Higgs particle under such conditions these alignment limits become very important.

The aim of this paper is to put some constraints on $ \textcolor{red}{m_{H^\pm}}$ - tan$\beta $ space in 2HDM by comparing CMS values of $\sigma_{H^\pm} \times BR(H^\pm \rightarrow tb^{\textcolor{red}{\pm}})$ with the computational values. Where  $\sigma_{H^\pm}$ stands for charged Higgs production cross section and $BR(H^\pm \rightarrow tb)$ stands for charged Higgs branching ratio into $tb^\mp$ channels. These constraints (experimental constraints) will be applied on charged Higgs parameter space in 2HDM both at 13 TeV and 8 TeV (\textcolor{red}{center of mass}) energies. Due to these constraints the $H^\pm$ parameter space in 2HDM will be squeezed which will reduce our labor work in finding charged Higgs. We also compare the expected and observed excluded regions of $H^\pm$ parameter space in 2HDM at both of these energies.\\

\section {Two Higgs Doublet Model}
SM has many limitations like dark matter, dark energy, \textcolor{red}{supersymmetry}, neutrino oscillation, baryon asymmetry etc, to fulfill these limitations we look beyond SM. 2HDM  is one of such beyond Standard Models. 2HDM is the simple extension of the Higgs sector of S.M.
There are a lot of motivations for the use of 2HDM , but the best one is \textcolor{red}{supersymmetry}. According to \textcolor{red}{supersymmetric} theories scalar particles are related to chiral multiplets while their complex conjugates are related to multiplets whose chiralities are opposite. Now as multiplets having opposite chiralities can not interact or couple together so only one Higgs doublet can not give mass to u like quarks (u, c, t quarks) and d like quarks (d, s, b quarks) simultaneously, in this regard we must have two Higgs doublets ($\phi_1$ and $\phi_2$), one will give mass to up like quarks (u like) and the second will give mass to down like quarks (d like) \cite{b1,b2}.
Another motivation for the use of 2HDM lies in the axion model. According to Peccie and Quenn there is a small CP violating term in Quantum Chromo dynamics (QCD) Lagrangian. Now if this Lagrangian has global U(1) symmetry then this CP violating term can be rotated away. Where the Lagrangian can have this type of symmetry only if it has two Higgs doublets. With only one Higgs doublet the Lagrangian is unable to have global U(1) symmetry and hence the small CP violating term in QCD Lagrangian can not be rotated away \cite{b3,b4}.\\
	
\subsection{Types of 2HDM}
On the basis of coupling of quarks and leptons to $\phi_1$ and $\phi_2$ the 2HDM is divided into four types. These types are; type I, type II, type X and type Y. Each one is described as ;
In type I u type or up like quarks (u, c and t quarks) and down like quarks (down, s and b quarks) both couple to $\phi_2$. 
In type II u like quarks couple to $\phi_2$ and d like quarks couple to $\phi_1$. One thing must be kept in mind that in type I and type II 2HDM, leptons only interact with that Higgs doublet, with which down like quarks interact. For instance here in type I down like quarks interact with second doublet $\phi_2$. So the lepton also interacts with $\phi_2$ in type I. Similarly in type II down like quarks interact with first doublet $\phi_1$ so the same is done by leptons too.
In  type X all up like quarks and down like quarks interact with $\phi_2$ while the leptons couple / interact with $\phi_1$. Here the coupling of leptons is specific / different from the couplings of u and d like quarks that is why such type of 2HDM is also called Lepton Specific 
In type Y up like quarks interact with $\phi_2$, down like quarks interact with $\phi_1$ but in contrast to type I and type II where all leptons interact with that Higgs doublet with which down type quarks interact, in type Y leptons interact to that Higgs doublet to which up type quarks couple, that is why this type of 2HDM is also called, Flipped.
The details of all types of 2HDM, in compact form are given in table \ref{tab1}.
As in type II 2HDM up like quarks interact with $\phi_2$, down like quarks and charged leptons interact with $\phi_1$. These interactions are called, Yukawa interactions, which are similar to the interactions of Higgs doublets with quarks and leptons in Minimal Super Symmetric Model (MSSM). That is why MSSM is a special type of 2HDM having same Yukawa interactions as that of type II 2HDM.
\end{multicols}
\begin{table}[h]
\centering
\begin{tabular}{|c|c|c|c|} \hline
Type & u like quarks & d like quarks & charged leptons\\ \hline
Type I & $\phi_2$ & $\phi_2$ & $\phi_2$\\ \hline
Type II & $\phi_2$ & $\phi_1$ & $\phi_1$\\ \hline
Type X & $\phi_2$ & $\phi_2$ & $\phi_1$\\ \hline
Type Y & $\phi_2$ & $\phi_1$ & $\phi_2$\\ \hline
\end{tabular}
\caption{Types of 2HDM and couplings of quarks and leptons to $\phi_1$ and $\phi_2$}
\label{tab1}
\end{table}
\begin{multicols}{2}

\subsection{Yukawa Interactions}		
For all types of 2HDM except FCNC  the most general Yukawa Interactions of up like, down like quarks and charged leptons with $H^\pm$ is given by the following equation \ref{eq1}. \cite{b5}
\begin{equation}
\label{eq1}
\begin{split}
L_{H^{\pm}}=-H^{+}[\frac{\sqrt{2}}{\nu}(m_uP_lX+m_dP_RY)V_{ud}\bar{u}d+ \\ \frac{\sqrt{2}}{\nu}m_lZ\bar{\nu}_ll_R]+H.C 
\end{split}
\end{equation}

Where $m_u$, $m_d$ and $m_l$ denote the masses of up like quarks , down like quarks and charged leptons respectively. H.C stands for Hermitian Conjugate, i.e. the Hermitian conjugate of the terms $-H^{+} [ \frac{\sqrt{2}}{\nu}  (m_u P_l X + m_d P_R Y) V_{ud} \bar{u}d + \frac{\sqrt{2}}{\nu} m_l Z\bar{\nu}_l l_R ] $, X, Y and Z represent the interactions of up like quarks, down like quarks and charged leptons with $H^{\pm}$ respectively whose values are different for different types of 2HDM and are given in table \ref{tab2}.

In equation \ref{eq1}, $v_{ud}$ represents an element of CKM matrix. The mod square of every member of the CKM matrix gives us the transition probability / conversion probability of one quark into another quark. For instance $|V_{ud}|^2$ represents the conversion probability of down quark into up quark. $|V_{us}|^2$  represents the conversion probability of strange quark into up quark. Generally  $|V_{ij}|^2$  represents the conversion probability of $j$ quark into $i$ quark.
\end{multicols}	 
\begin{table}[h]
\centering
\begin{tabular}{|c|c|c|c|} \hline
Type & u like quarks (X) & d like quarks (Y) & charged leptons (Z)\\ \hline
Type I & $\cot{\beta}$ & $\cot{\beta}$ & $\cot{\beta}$\\ \hline
Type II & $\cot{\beta}$ & -$\tan{\beta}$ & -$\tan{\beta}$\\ \hline
Type X & $\cot{\beta}$ & $\cot{\beta}$ & -$\tan{\beta}$\\ \hline
Type Y & $\cot{\beta}$ & -$\tan{\beta}$ & $\cot{\beta}$\\ \hline
\end{tabular}
\caption{X,Y \& Z values in 2HDMs}
\label{tab2}
\end{table}
\begin{multicols}{2}
				
\subsection{The Higgs Potential}
Every field has \textcolor{red}{its} own potential, like Gravitational field has \textcolor{red}{its} own potential which depends upon the masses of bodies and the separation distance between them. The Electric field has \textcolor{red}{its} own potential which depends upon the charges and the separation distance between them. In the same manner the Higgs field has \textcolor{red}{its} own potential called Higgs potential ($V_H$) which depends on $\phi_1$  and $\phi_2$  and some other mixing parameters. $V_H$ in most simplified form is given in equation \ref{eq2} \cite{b5}.

\begin{equation}	
\label{eq2}
\begin{split}
V_H = m_{11}^2\phi_{I}^\dagger\phi_I+m_{22}^2\phi_{II}^\dagger\phi_{II}-[m_{12}^2\phi_{I}^\dagger\phi_{II}+H.C] + \\ \frac{\lambda_1}{2}(\phi_I^\dagger\phi_I)^2+\frac{\lambda_2}{2}(\phi_{II}^\dagger\phi_{II})^2+\lambda_3(\phi_I^\dagger\phi_I)(\phi_{II}^\dagger\phi_{II}) + \\
\lambda_4(\phi_I^\dagger\phi_{II})(\phi_{II}^\dagger\phi_I)+[\frac{\lambda_5}{2}(\phi_I^\dagger\phi_{II})^2+ \lambda_6(\phi_I^\dagger\phi_I)(\phi_I^\dagger\phi_{II})+ \\ \lambda_7(\phi_{II}^\dagger\phi_{II})(\phi_I^\dagger\phi_{II})+H.C]
\end{split}
\end{equation}
	
Where $\phi_{I}$ and $\phi_{II}$ denote $\phi_1$ and $\phi_2$ respectively. The first H.C represents the Hermitian conjugate of the term, $m_{12}^2\phi_I^{\dagger} \phi_{II}$, and the last H.C gives the Hermitian conjugate of the three terms enclosed in the last square bracket. $m^{2}_{11,22,12}$ are called mass mixing parameters. Here the parameters $m^2_{11,22}$, $\lambda_{1,-,4}$ are real and $m^2_{12}$, $\lambda_{5,-,7}$ are in general complex. So the Higgs potential given in equation 2 depends on six real parameters and \textcolor{red}{four} complex parameters.
\subsection{$Z_2$ and CP Symmetries}

To skip FCNC safely, we impose $Z_2$ symmetry on Higgs Potential given in equation \ref{eq2}. $Z_2$ symmetry says; $\phi_1 \rightarrow \phi_1$ and $\phi_2 \rightarrow - \phi_2$\\  

Now Higgs potential is said to obey the above conditions of $Z_2$ symmetry if $\lambda_6 = \lambda_7 = m_{12}^2 = 0$ \cite{b6} in equation \ref{eq2}. The general 2HDM also permits the CP violation which can be overcome by imposing $Z_2$ symmetry on Higgs Potential \cite{b7,b8,b9}. Let for instance we assume that $\lambda_5$ and $m_{12}^2$ are real by assuming that the CP symmetry is invariant under such assumption. Also we assume that, $\lambda_6 = \lambda_7 = 0$ but $m_{12}^2 \neq 0$, under such conditions $Z_2$ symmetry "softly breaks" which otherwise breaks if $\lambda_6=\lambda_7 = m_{12}^2 = 0$ \cite{b10,b11,b12,b13}. Under such assumptions, then equation \ref{eq2} can be written, as given in equation \ref{eq3}.

\begin{equation}
\label{eq3}	
\begin{split}
V_H=m_{11}^2\phi_I^\dagger\phi_I+m_{22}^2\phi_{II}^\dagger\phi_{II}-[m_{12}^2\phi_I^\dagger\phi_{II}+H.C]+ \\ \frac{\lambda_1}{2}(\phi_I^\dagger\phi_I)^2+\frac{\lambda_2}{2}(\phi_{II}^\dagger\phi_{II})^2 + \lambda_3(\phi_I^\dagger\phi_I)(\phi_{II}^\dagger\phi_{II})+ \\ \lambda_4(\phi_I^\dagger\phi_{II})(\phi_{II}^\dagger\phi_I)+[\frac{\lambda_5}{2}(\phi_I^\dagger\phi_{II})^2+H.C]
\end{split}
\end{equation}
	
\section{Computer Simulations}
The main purpose of our paper is to put some upper limits on $H^\pm$ parameter space in 2HDM by using the computational values of  $\sigma^\pm \times BR(H^\pm)$ and then comparing with the CMS results. Therefore it is necessary first to calculate the branching fractions / branching ratios of charged Higgs into each of the fermionic channels. We calculate these branching ratios by using a software called "Two Higgs Doublet Model Calculator"  abbreviated as 2HDMC \cite{b14}. There are a lot of versions of 2HDMC ,we use 1.7.0 version of 2HDMC for our calculations. 
	In 2HDMC we can carry out our calculations of branching fractions in various basis and each of these basis depends on different parameter space of 2HDM. These basis and the parameters on which these basis depend are listed below;
One of the best choices is to pick the physical mass basis denoted by "CalcPhys" in 2HDMC. CalcPhys depends on; $m_h$, $m_H$, $m_A$, $\textcolor{red}{m_{H^\pm}}$, alignment limit $\sin{(\beta-\alpha)}$, $\lambda_6, \lambda_7$, mass mixing parameter $(m_{12}^2)$ and ratio of two vacuum expectation values $(\tan{\beta})$. For our calculations we use physical mass basis ie, "CalcPhys". One basis is "CalcHMSSM" which depends on $m_h , m_A$ and $\tan{\beta}$. One basis is hybrid basis denoted by "CalcHybrid" and depends on ; $m_h, m_H, \cos{(\beta-\alpha)}, Z_4, Z_5, Z_7$ and $\tan{\beta}$. One basis is Higgs basis denoted by "CalcHiggs" which depends on; $\lambda_{1,2,3,4,5,6,7}$ and $\textcolor{red}{m_{H^\pm}}$. Along with these there are some other basis like; "CalcH" and "CalcMSSM" which depend on different parameters.
	Before putting the values of parameters in 2HDMC we need to do some assumptions, through out this paper, we assume, $\textcolor{red}{m_{H^\pm}} = m_h = m_A$ (mass degeneracy) to avoid charged Higgs decay into gauge bosons ie, to avoid  $H^\pm \to W^\pm \phi$ where $(\phi = H , A)$ so that the decay of $H^\pm$ into $W^\pm \phi$ will be kinematically forbidden as  $W^\pm,\phi$ have greater mass than $H^\pm$. Moreover $\sin{(\beta-\alpha)}$ is assumed to be equal to unity because in this alignment limit the CP even light Higgs (h) behaves like SM Higgs. When  Higgs masses are assumed to be degenerate then the mass mixing parameter $(m_{12}^2 )$ can be found by the formula;

\begin{equation}
\label{eq4}
m_{12}^2= m_A^2 \sin{\beta} \cos{\beta}
\end{equation}  
	
\begin{figure}[!tbp]
  \centering
  \begin{minipage}[b]{0.4\textwidth}
    \includegraphics[width=8cm]{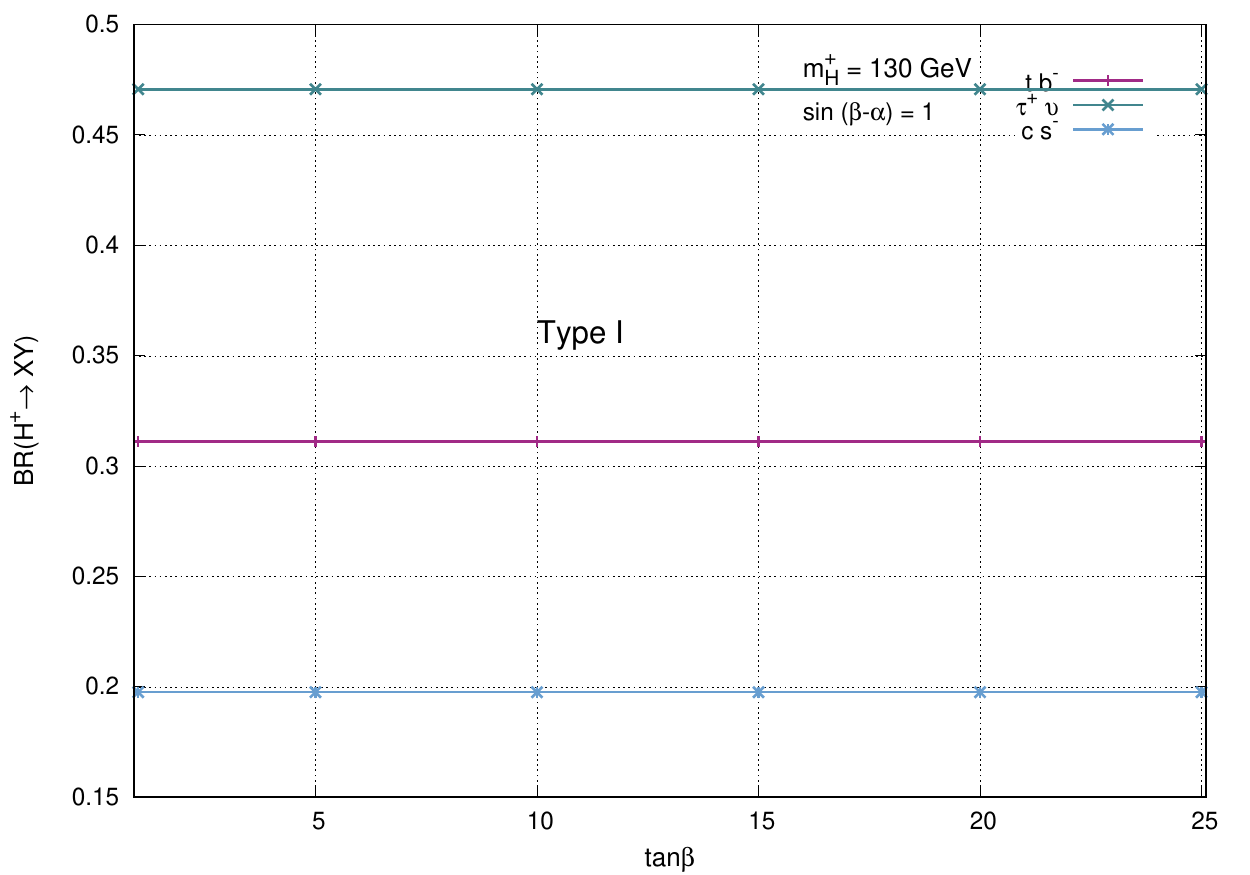}
    \caption{The plot is shown between tan$\beta$ and BR ($H^+$)}
    \label{fig1}
  \end{minipage}
  \qquad
  \begin{minipage}[b]{0.4\textwidth}
    \includegraphics[width=8cm]{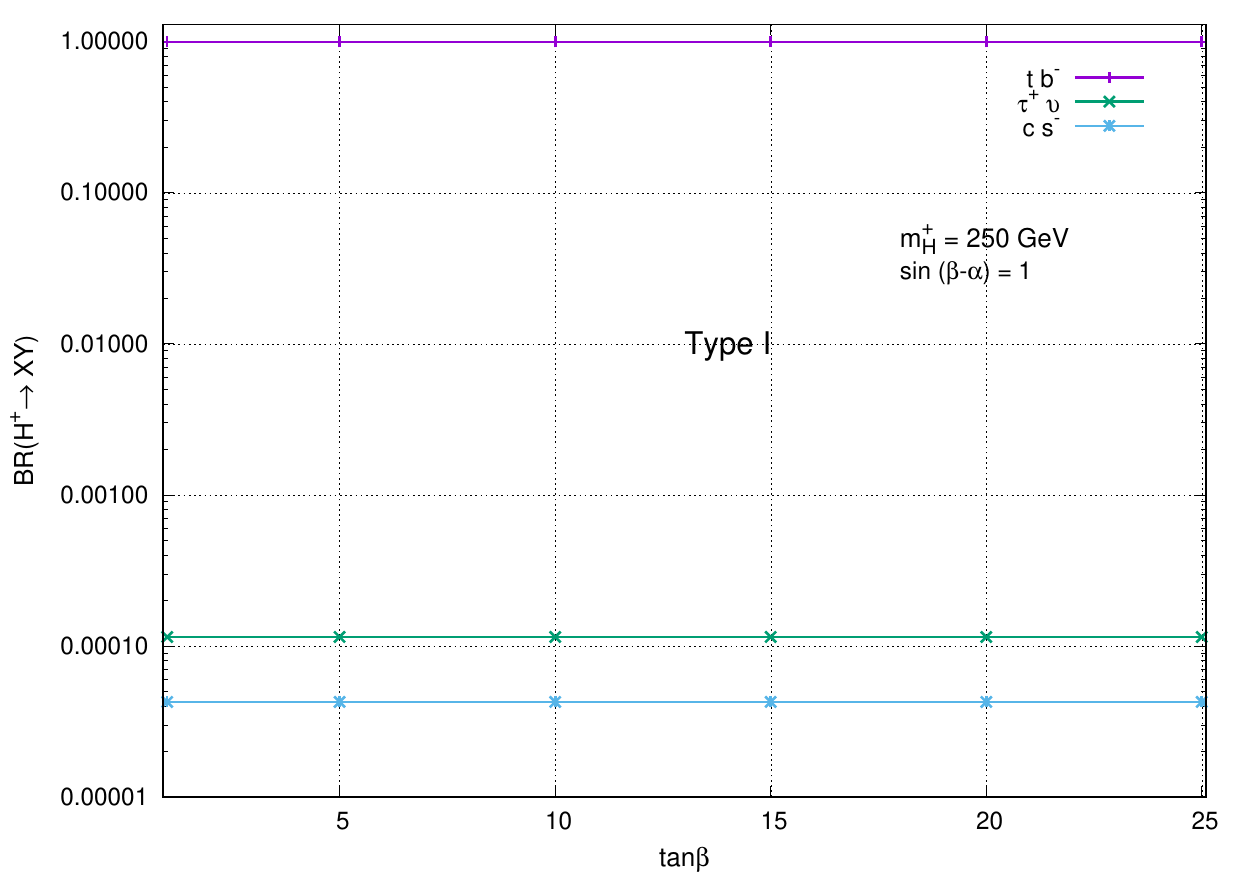}
    \caption{The plot between tan$\beta$ and BR ($H^+$)}
    \label{fig2}
  \end{minipage}
    \qquad
  \begin{minipage}[b]{0.4\textwidth}
    \includegraphics[width=8cm]{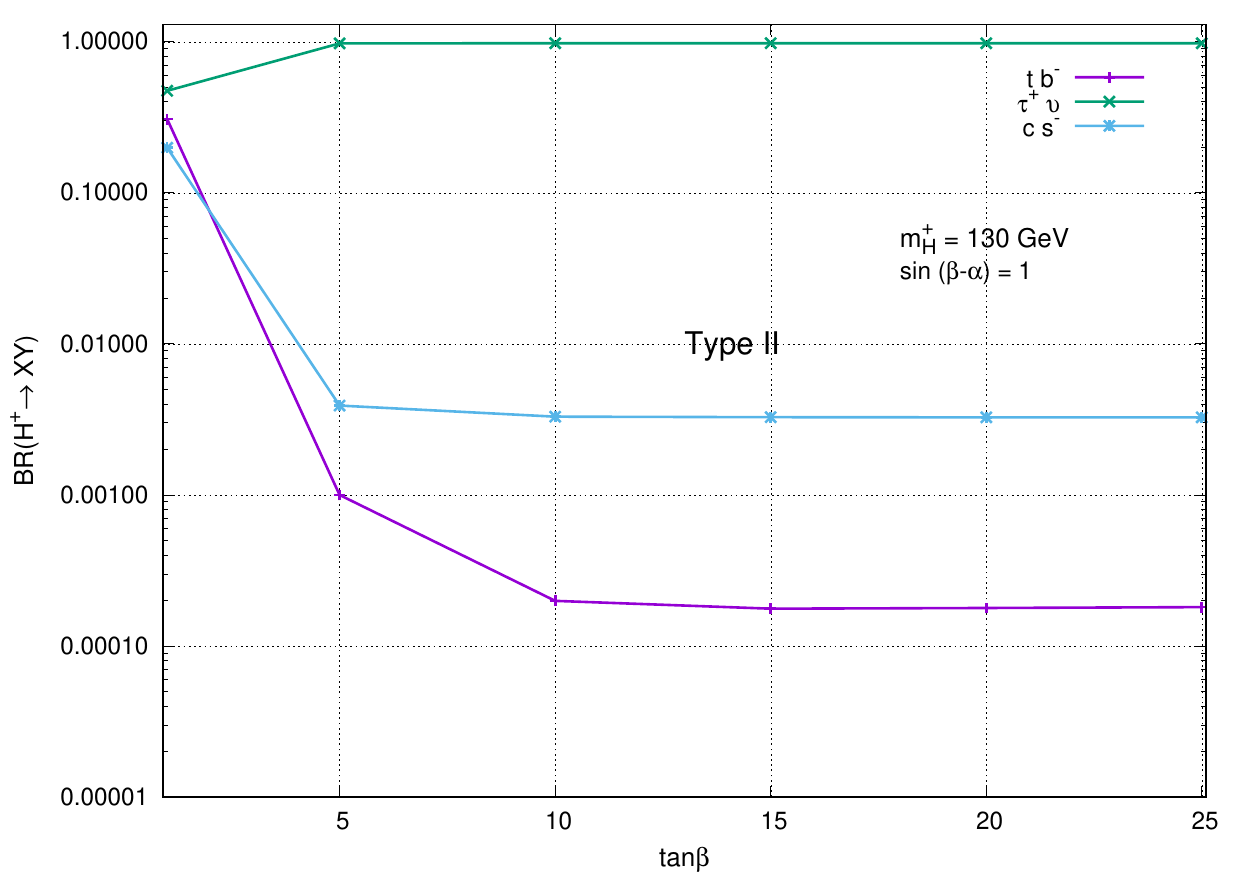}
    \caption{The plot between tan$\beta$ and BR ($H^+$)}
    \label{fig3}
  \end{minipage}
      \qquad
  \begin{minipage}[b]{0.4\textwidth}
    \includegraphics[width=8cm]{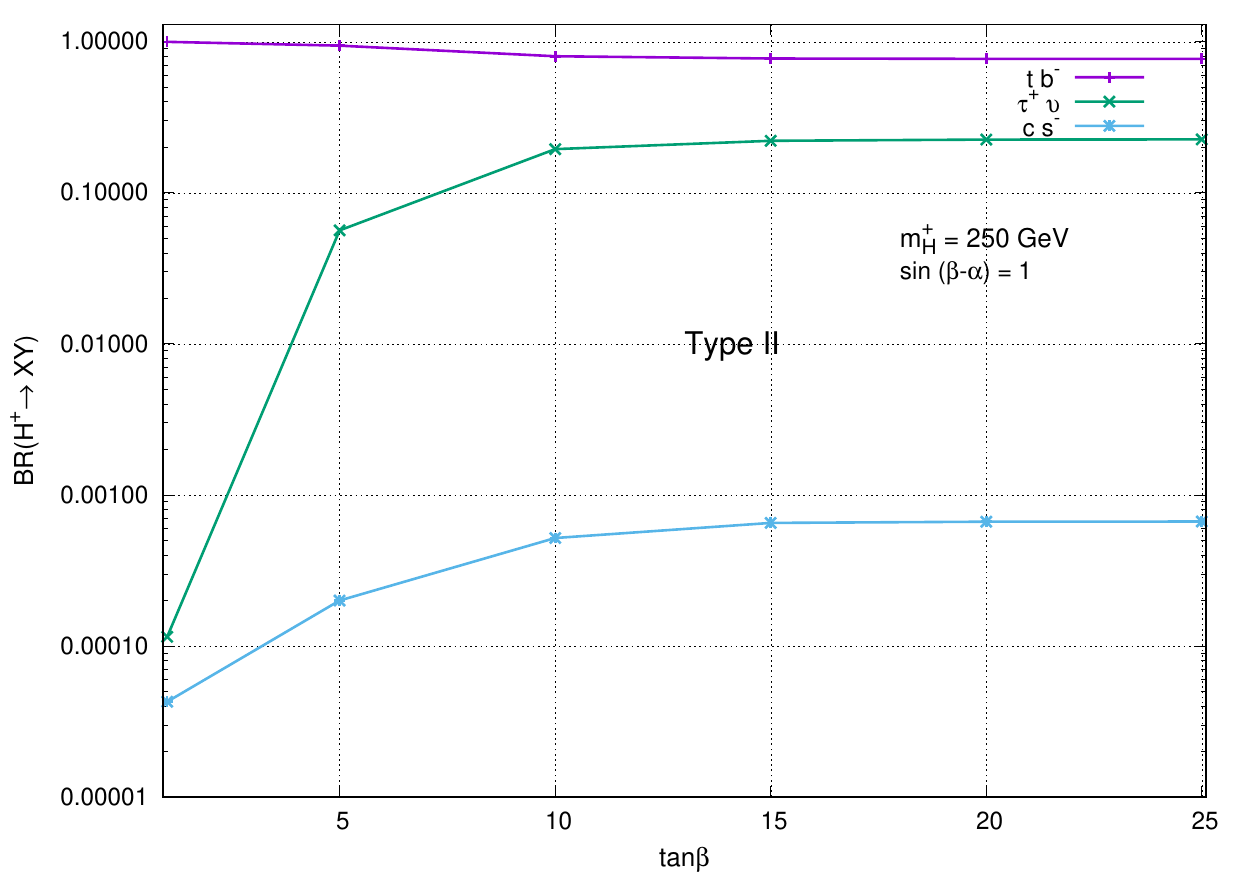}
    \caption{The plot between tan$\beta$ and BR ($H^+$)}
    \label{fig4}
  \end{minipage}
      \qquad
  \begin{minipage}[b]{0.4\textwidth}
    \includegraphics[width=8cm]{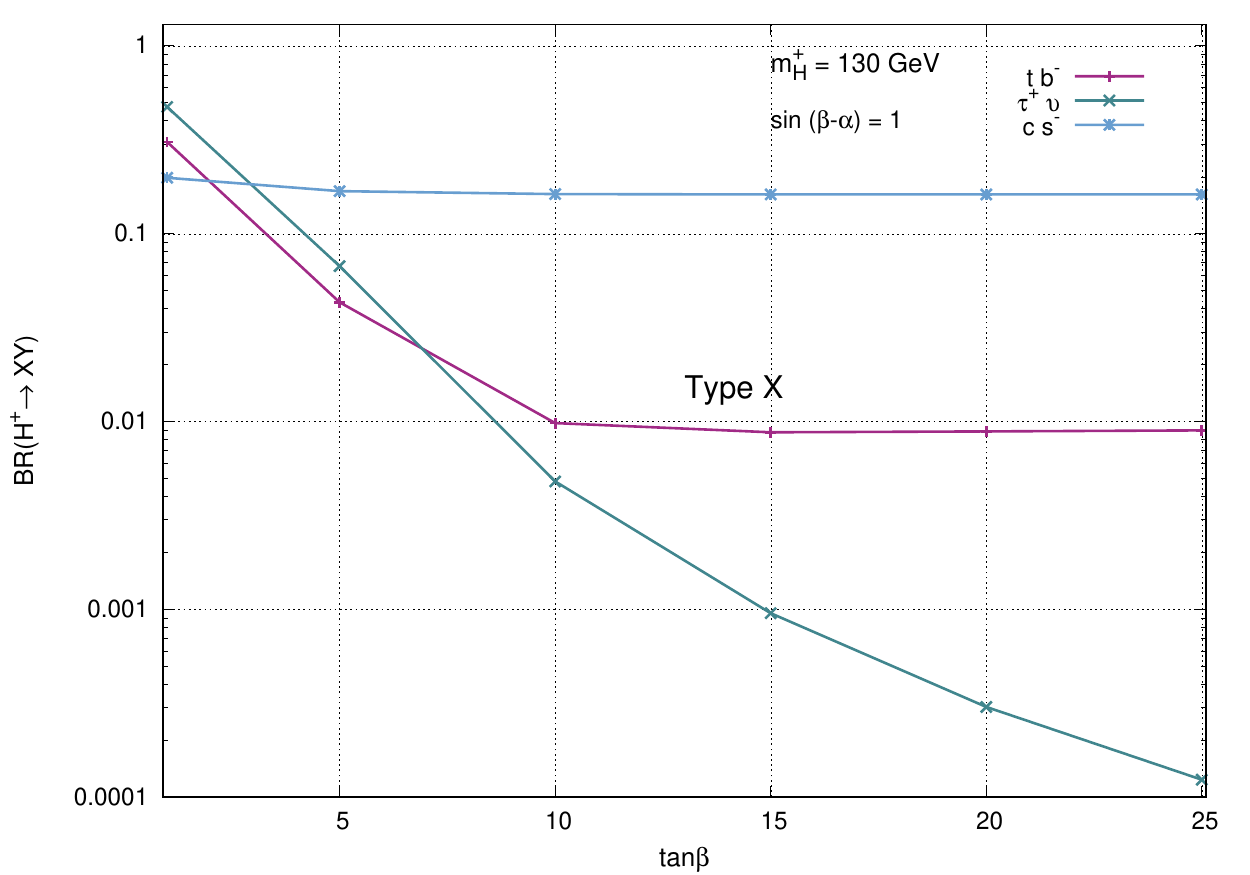}
    \caption{The plot between tan$\beta$ and BR ($H^+$)}
    \label{fig5}
  \end{minipage}
      \qquad
  \begin{minipage}[b]{0.4\textwidth}
    \includegraphics[width=8cm]{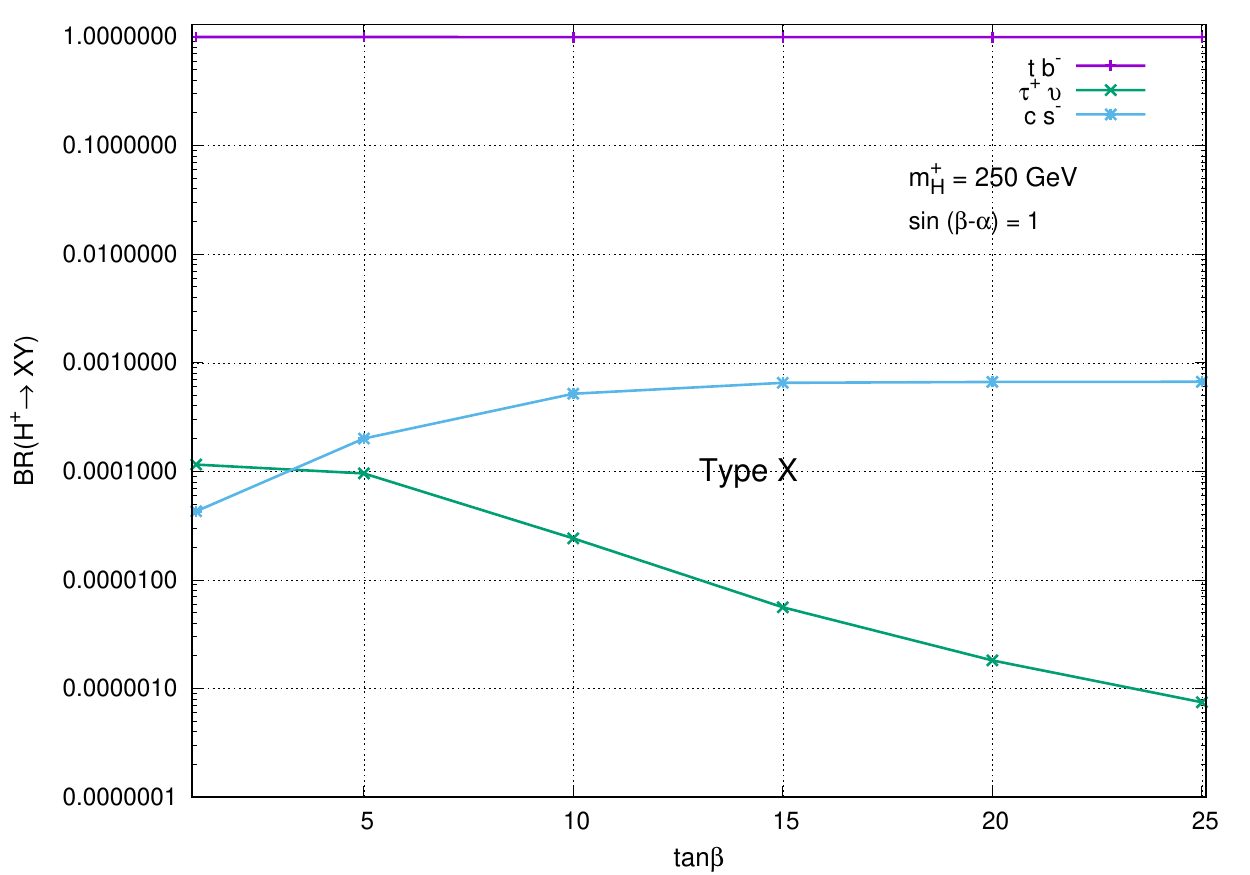}
    \caption{plot between tan$\beta$ and BR ($H^+$)}
    \label{fig6}
  \end{minipage}
\end{figure}

\begin{figure}[!tbp]
  \centering      
  \begin{minipage}[b]{0.4\textwidth}
    \includegraphics[width=8cm]{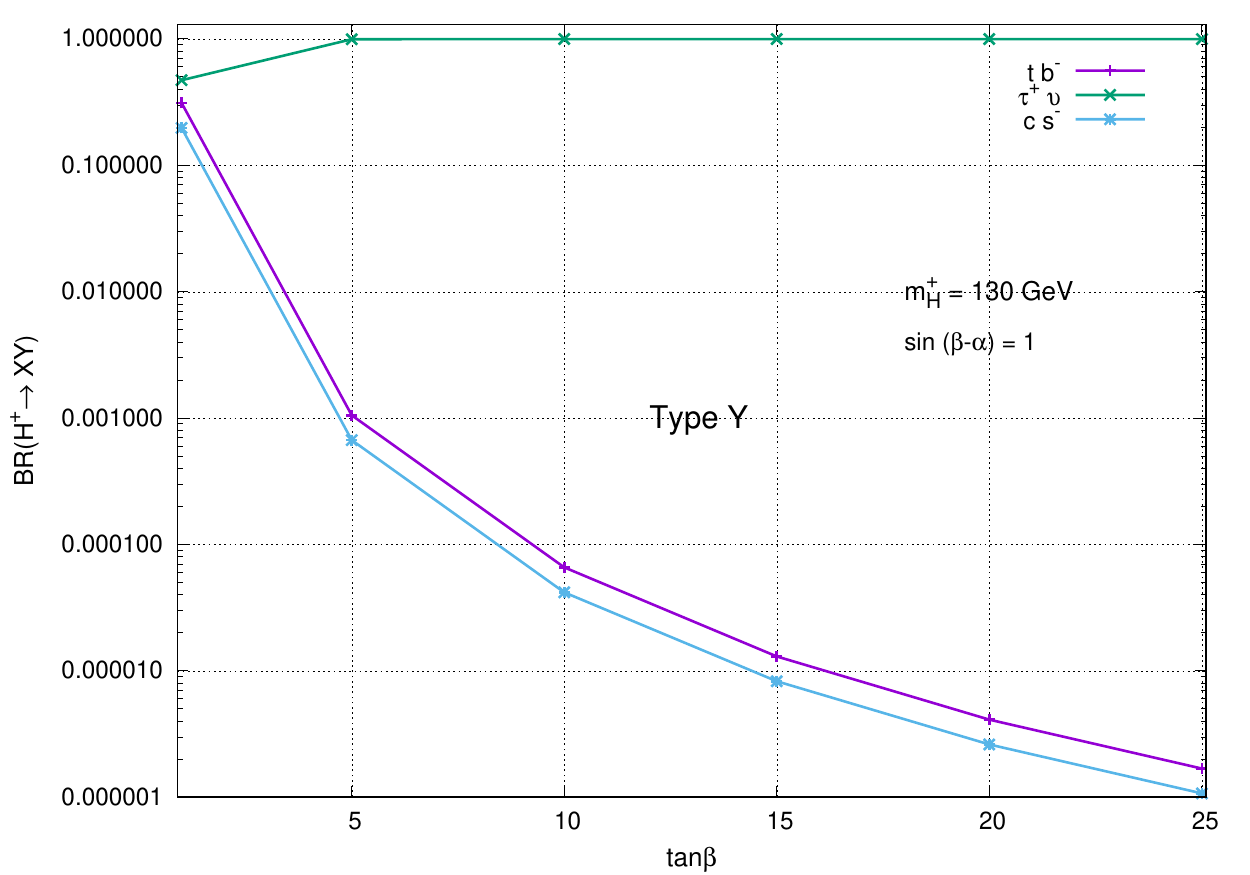}
    \caption{The plot between tan$\beta$ and BR ($H^+$)}
    \label{fig7}
  \end{minipage}
      \qquad
  \begin{minipage}[b]{0.4\textwidth}
    \includegraphics[width=8cm]{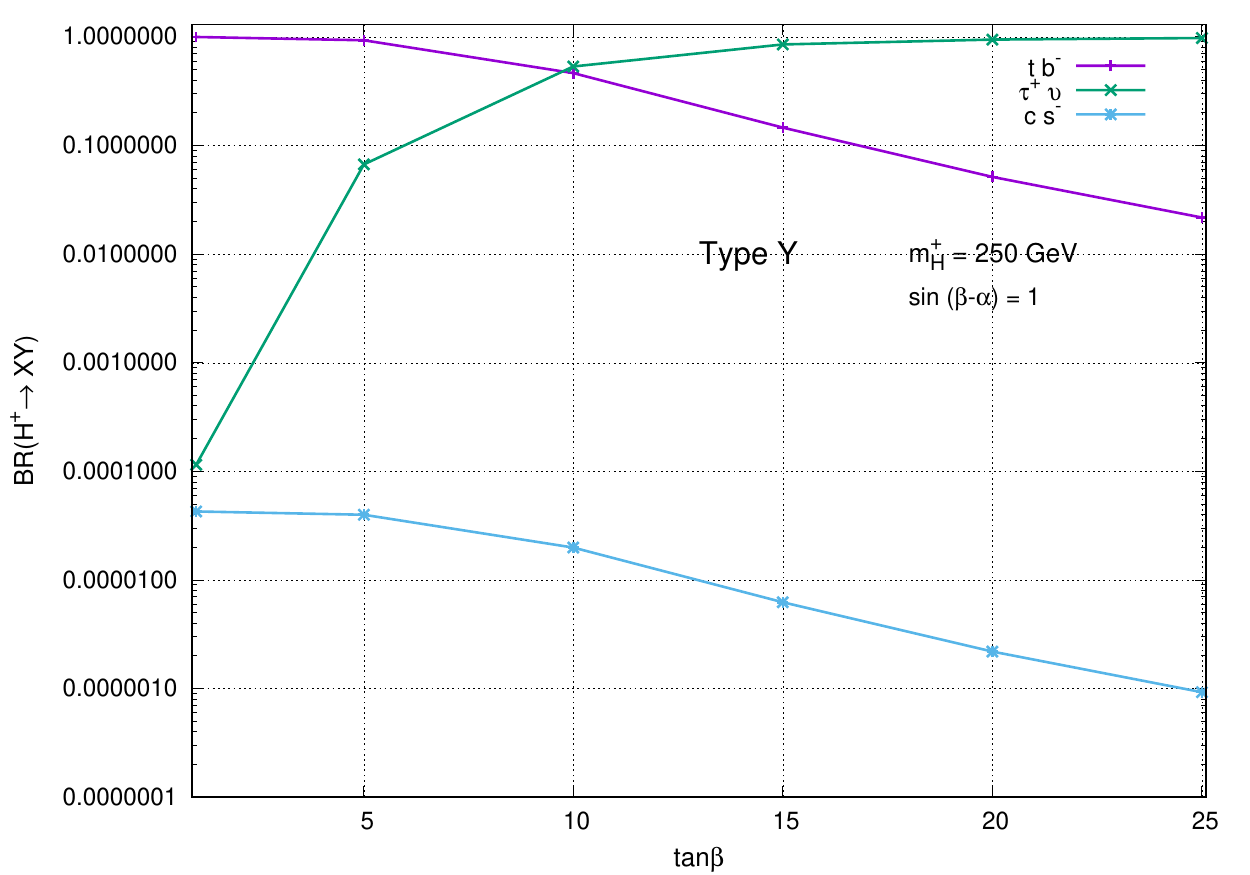}
    \caption{The plot between tan$\beta$ and BR ($H^+$)}
    \label{fig8}
  \end{minipage}
\end{figure}

Our main decay modes / channels of $H^\pm$ are; top bottom quarks (tb), tau nu leptons $(\tau,\nu)$ and charm strange quarks (CS). There are a lot of other decay modes too of charged Higgs but they are highly suppressed by these three decay modes and hence leaving the remaining decay modes non significant. Now lets calculate the branching fractions of charged Higgs into these three decay modes / channels for all types of 2HDM for both light and heavy $H^\pm$ scenarios, these branching fractions are given in the tables from Table \ref{tab3} to Table \ref{tab10} and the corresponding plots are given in the figures from \textcolor{red}{Fig.\ref{fig1}} to Fig.\ref{fig8} respectively.\\
\end{multicols}
\begin{table}[ht]
\centering
\begin{tabular}{|c|c|c|c|c|} \hline
S.NO. &  $\tan{\beta}$ & BR($H^+\to t b^-$) & BR($H^+\to \tau^+ \nu$) & BR($H^+\to c s^-$) \\ \hline
1 & 1 & 3.1 $\times$ ${10}^{-1}$ & 4.7 $\times$ ${10}^{-1}$ & 2.0 $\times$ ${{10}^{-1}}$\\ \hline
2 & 5 & 3.1 $\times$ ${{10}^{-1}}$ & 4.7 $\times$ ${{10}^{-1}}$ & 2.0 $\times$ ${{10}^{-1}}$ \\ \hline
3 &	10 & 3.1 $\times$ ${{10}^{-1}}$ & 4.7 $\times$ ${{10}^{-1}}$ & 2.0 $\times$ ${{10}^{-1}}$ \\ \hline
4 &	15 & 3.1 $\times$ ${{10}^{-1}}$ & 4.7 $\times$ ${{10}^{-1}}$ & 2.0 $\times$ ${{10}^{-1}}$ \\ \hline
5 & 20 & 3.1 $\times$ ${{10}^{-1}}$ & 4.7 $\times$ ${{10}^{-1}}$ & 2.0 $\times$ ${{10}^{-1}}$ \\ \hline
6 & 25 & 3.1 $\times$ ${{10}^{-1}}$ & 4.7 $\times$ ${{10}^{-1}}$ & 2.0 $\times$ ${{10}^{-1}}$ \\ \hline
\end{tabular}
\caption{The charged Higgs Branching Ratios with respect to different values of tan$\beta$ calculated by 2HDMC in type I for Light $H^+$ scenario.}
\label{tab3}
\end{table}
	
\begin{table}[ht]
\centering
\begin{tabular}{|c|c|c|c|c|} \hline
S.NO. &  tan$\beta$ & BR($H^+\to t b^-$) & BR($H^+\to \tau^+ \nu$) & BR($H^+\to c s^-$) \\ \hline
1 & 1 & 10.0 $\times$ ${{10}^{-1}}$ & 1.1 $\times$ ${{10}^{-4}}$ & 4.3 $\times$ ${{10}^{-5}}$\\ \hline
2 & 5 & 10.0 $\times$ ${{10}^{-1}}$ & 1.1 $\times$ ${{10}^{-4}}$ & 4.3 $\times$ ${{10}^{-5}}$\\ \hline
3 & 10 & 10.0 $\times$ ${{10}^{-1}}$ & 1.1 $\times$ ${{10}^{-4}}$ & 4.3 $\times$ ${{10}^{-5}}$\\ \hline
4 & 15 & 10.0 $\times$ ${{10}^{-1}}$ & 1.1 $\times$ ${{10}^{-4}}$ & 4.3 $\times$ ${{10}^{-5}}$\\ \hline
5 & 20 & 10.0 $\times$ ${{10}^{-1}}$ & 1.1 $\times$ ${{10}^{-4}}$ & 4.3 $\times$ ${{10}^{-5}}$\\ \hline
6 & 25 & 10.0 $\times$ ${{10}^{-1}}$ & 1.1 $\times$ ${{10}^{-4}}$ & 4.3 $\times$ ${{10}^{-5}}$\\ \hline
\end{tabular}
\caption{The charged Higgs Branching Ratios with respect to different values of tan$\beta$ calculated by 2HDMC in type I for heavy $H^+$ scenario.}
\label{tab4}
\end{table}

\begin{table}[ht]
\centering
\begin{tabular}{|c|c|c|c|c|} \hline
S.NO. &  tan$\beta$ & BR($H^+\to t b^-$) & BR($H^+\to \tau^+ \nu$) & BR($H^+\to c s^-$) \\ \hline
1 & 1 & 3.1 $\times$ ${{10}^{-1}}$ & 4.7 $\times$ ${{10}^{-1}}$ & 2.0 $\times$ ${{10}^{-1}}$\\ \hline
2 &	5 & 1.0 $\times$ ${{10}^{-3}}$ & 9.7 $\times$ ${{10}^{-1}}$ & 4.0 $\times$ ${{10}^{-3}}$\\ \hline
3 &	10 & 2.0 $\times$ ${{10}^{-4}}$ & 9.8 $\times$ ${{10}^{-1}}$ &	3.3 $\times$ ${{10}^{-3}}$\\ \hline
4 & 15 & 1.8 $\times$ ${{10}^{-4}}$ & 9.8 $\times$ ${{10}^{-1}}$ &	3.3 $\times$ ${{10}^{-3}}$\\ \hline
5 &	20 & 1.8 $\times$ ${{10}^{-4}}$ & 9.8 $\times$ ${{10}^{-1}}$ & 3.3 $\times$ ${{10}^{-3}}$\\ \hline	
6 & 25 & 1.8 $\times$ ${{10}^{-4}}$ & 9.8 $\times$ ${{10}^{-1}}$ & 3.3 $\times$ ${{10}^{-3}}$\\ \hline
\end{tabular}
\caption{The charged Higgs Branching Ratios with respect to different values of tan$\beta$ calculated by 2HDMC in type II for light $H^+$ scenario.}
\label{tab5}
\end{table}

\begin{table}[ht]
\centering
\begin{tabular}{|l|l|l|l|l|} \hline
S.NO. &  tan$\beta$ & BR($H^+\to t b^-$) & BR($H^+\to \tau^+ \nu$) & BR($H^+\to c s^-$) \\ \hline
1 & 1 & 10.0 $\times$ ${{10}^{-1}}$ & 1.2 $\times$ ${{10}^{-4}}$ & 4.3 $\times$ ${{10}^{-5}}$\\ \hline
2 & 5 & 9.4 $\times$ ${{10}^{-1}}$ & 5.6 $\times$ ${{10}^{-2}}$ & 2.0 $\times$ ${{10}^{-4}}$ \\ \hline
3 & 10 & 8.0 $\times$ ${{10}^{-1}}$ & 1.9 $\times$ ${{10}^{-1}}$ & 5.2 $\times$ ${{10}^{-4}}$\\ \hline
4 & 15 & 7.7 $\times$ ${{10}^{-1}}$ & 2.2 $\times$ ${{10}^{-1}}$ & 6.5 $\times$ ${{10}^{-4}}$ \\ \hline
5 & 20 & 7.7 $\times$ ${{10}^{-1}}$ & 2.2 $\times$ ${{10}^{-1}}$ & 6.7 $\times$ ${{10}^{-4}}$\\ \hline
6 & 25 & 7.7 $\times$ ${{10}^{-1}}$ & 2.3 $\times$ ${{10}^{-1}}$ & 6.7 $\times$ ${{10}^{-4}}$\\ \hline
\end{tabular}
\caption{The charged Higgs Branching Ratios with respect to different values of tan$\beta$ calculated by 2HDMC in type II for heavy $H^+$ scenario.}
\label{tab6}
\end{table}

\begin{table}[ht]
\centering
\begin{tabular}{|c|c|c|c|c|} \hline
S.NO. &  tan$\beta$ & BR($H^+\to t b^-$) & BR($H^+\to \tau^+ \nu$) & BR($H^+\to c s^-$) \\ \hline
1 & 1 & 3.1 $\times$ ${{10}^{-1}}$ & 4.7 $\times$ ${{10}^{-1}}$ & 2.0 $\times$ ${{10}^{-1}}$ \\ \hline
2 & 5 & 4.3 $\times$ ${{10}^{-2}}$ & 6.7 $\times$ ${{10}^{-2}}$ & 1.7 $\times$ ${{10}^{-1}}$ \\ \hline
3 & 10 & 9.8 $\times$ ${{10}^{-3}}$ & 4.8 $\times$ ${{10}^{-3}}$ & 1.6 $\times$ ${{10}^{-1}}$ \\ \hline
4 & 15 & 8.8 $\times$ ${{10}^{-3}}$ & 9.6 $\times$ ${{10}^{-4}}$ & 1.6 $\times$ ${{10}^{-1}}$ \\ \hline
5 & 20 & 8.8 $\times$ ${{10}^{-3}}$ & 3.0 $\times$ ${{10}^{-4}}$ & 1.6 $\times$ ${{10}^{-1}}$ \\ \hline
6 & 25 & 9.0 $\times$ ${{10}^{-3}}$ & 1.2 $\times$ ${{10}^{-4}}$ & 1.6 $\times$ ${{10}^{-1}}$\\ \hline
\end{tabular}
\caption{The charged Higgs Branching Ratios with respect to different values of tan$\beta$ calculated by 2HDMC in type X for light $H^+$ scenario.}
\label{tab7}
\end{table}

\begin{table}[ht]
\centering
\begin{tabular}{|c|c|c|c|c|} \hline
S.NO. &  tan$\beta$ & BR($H^+\to t b^-$) & BR($H^+\to \tau^+ \nu$) & BR($H^+\to c s^-$) \\ \hline
1 & 1 & 10.0 $\times$ ${{10}^{-1}}$ & 1.2 $\times$ ${{10}^{-4}}$ & 4.3 $\times$ ${{10}^{-5}}$\\ \hline
2 & 5 & 10.0 $\times$ ${{10}^{-1}}$ & 9.6 $\times$ ${{10}^{-5}}$ & 2.1 $\times$ ${{10}^{-4}}$\\ \hline
3 & 10 & 10.0 $\times$ ${{10}^{-1}}$ & 2.4 $\times$ ${{10}^{-5}}$ & 7.2 $\times$ ${{10}^{-4}}$\\ \hline
4 & 15 & 9.9 $\times$ ${{10}^{-1}}$ & 5.6 $\times$ ${{10}^{-6}}$ & 8.4 $\times$ ${{10}^{-4}}$\\ \hline
5 & 20 & 9.9 $\times$ ${{10}^{-1}}$ & 1.8 $\times$ ${{10}^{-6}}$ & 8.6 $\times$ ${{10}^{-4}}$\\ \hline
6 & 25 & 9.9 $\times$ ${{10}^{-1}}$ & 7.5 $\times$ ${{10}^{-7}}$ & 8.6 $\times$ ${{10}^{-4}}$\\ \hline
\end{tabular}
\caption{The charged Higgs Branching Ratios with respect to different values of tan$\beta$ calculated by 2HDMC in type X for heavy $H^+$ scenario.}
\label{tab8}
\end{table}

\begin{table}[ht]
\centering
\begin{tabular}{|c|c|c|c|c|} \hline
S.NO. &  tan$\beta$ & BR($H^+\to t b^-$) & BR($H^+\to \tau^+ \nu$) & BR($H^+\to c s^-$) \\ \hline
1 & 1 & 3.1 $\times$ ${{10}^{-1}}$ & 4.7 $\times$ ${{10}^{-1}}$ & 2.0 $\times$ ${{10}^{-1}}$\\ \hline
2 &	5 &	1.0 $\times$ ${{10}^{-3}}$ & 9.9 $\times$ ${{10}^{-1}}$ & 6.7 $\times$ ${{10}^{-4}}$\\ \hline
3 & 10 & 6.6 $\times$ ${{10}^{-5}}$ & 10.0 $\times$ ${{10}^{-1}}$ & 4.2 $\times$ ${{10}^{-5}}$\\ \hline
4 & 15 & 1.3 $\times$ ${{10}^{-5}}$ & 10.0 $\times$ ${{10}^{-1}}$ & 8.3 $\times$ ${{10}^{-6}}$ \\ \hline
5 & 20 & 4.1 $\times$ ${{10}^{-6}}$ & 10.0 $\times$ ${{10}^{-1}}$ & 2.6 $\times$ ${{10}^{-6}}$ \\ \hline
6 &	25 & 1.7 $\times$ ${{10}^{-6}}$ & 10.0 $\times$ ${{10}^{-1}}$ & 1.1 $\times$ ${{10}^{-6}}$\\ \hline
\end{tabular}
\caption{The charged Higgs Branching Ratios with respect to different values of tan$\beta$ calculated by 2HDMC in type Y for light $H^+$ scenario.}
\label{tab9}
\end{table}

\begin{table}[!ht]
\centering
\begin{tabular}{|c|c|c|c|c|} \hline
S.NO. &  tan$\beta$ & BR($H^+\to t b^-$) & BR($H^+\to \tau^+ \nu$) & BR($H^+\to c s^-$) \\ \hline
1 & 1 & 10.0 $\times$ ${{10}^{-1}}$ & 1.1 $\times$ ${{10}^{-4}}$ & 4.3 $\times$ ${{10}^{-5}}$\\ \hline
2 & 5 & 9.3 $\times$ ${{10}^{-1}}$ & 6.7 $\times$ ${{10}^{-2}}$ & 4.0 $\times$ ${{10}^{-5}}$ \\ \hline
3 & 10 & 4.6 $\times$ ${{10}^{-1}}$ & 5.3 $\times$ ${{10}^{-1}}$ & 2.0 $\times$ ${{10}^{-5}}$ \\ \hline
4 & 15 & 1.5 $\times$ ${{10}^{-1}}$ & 8.5 $\times$ ${{10}^{-1}}$ & 6.2 $\times$ ${{10}^{-6}}$ \\ \hline
5 & 20 & 5.1 $\times$ ${{10}^{-2}}$ & 9.5 $\times$ ${{10}^{-1}}$ & 2.2 $\times$ ${{10}^{-6}}$ \\ \hline
6 & 25 & 2.2 $\times$ ${{10}^{-2} }$& 9.7 $\times$ ${{10}^{-1}}$ & 9.3 $\times$ ${{10}^{-7}}$\\ \hline
\end{tabular}
\caption{The charged Higgs Branching Ratios with respect to different values of tan$\beta$ calculated by 2HDMC in type Y for heavy $H^+$ scenario.}
\label{tab10}
\end{table}

\begin{multicols}{2}
From all of the tables and figures given in section 3 it is obvious that, for the given range of $\tan{\beta}$ and specially for $\tan{\beta} \leq 5$ the dominant decay channel for light scenario is $H^+\to \bar{\tau} \nu$ while for the heavy scenario,  $H^+\to t\bar{b} $ is the dominant decay channel \textcolor{red}{for $\tan{\beta} < 10$} while the others are greatly suppressed. That is why one can only consider  $H^+\to t\bar{b}$ channel \textcolor{red}{for $\tan{\beta} < 10$} as in this article are working in the heavy Higgs scenario.  

\section{Exclusion Regions of $\textcolor{red}{m_{H^\pm}}$ - tan$\beta$ Parameter Space  in 2HDM}
\begin{figure}[!tbp]
  \centering

  \begin{minipage}[b]{0.4\textwidth}
    \includegraphics[width=8cm]{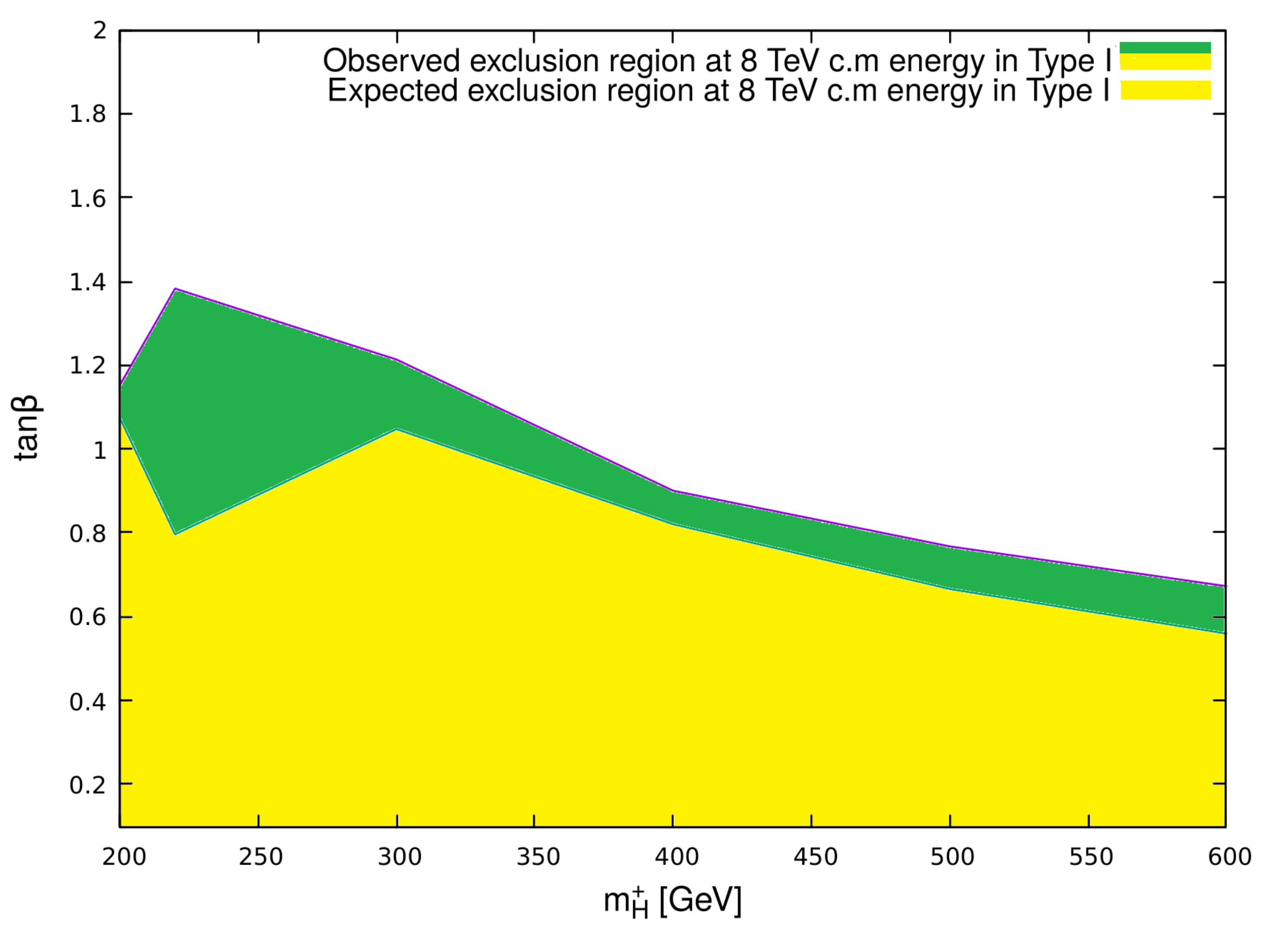}
\caption{The comparison of expected (yellow) and observed (green plus yellow) exclusion regions of charged Higgs parameter space in type1 2HDM at 8 TeV energy.}
    \label{fig9}
  \end{minipage}
      \qquad
 \begin{minipage}[b]{0.4\textwidth}
    \includegraphics[width=8cm]{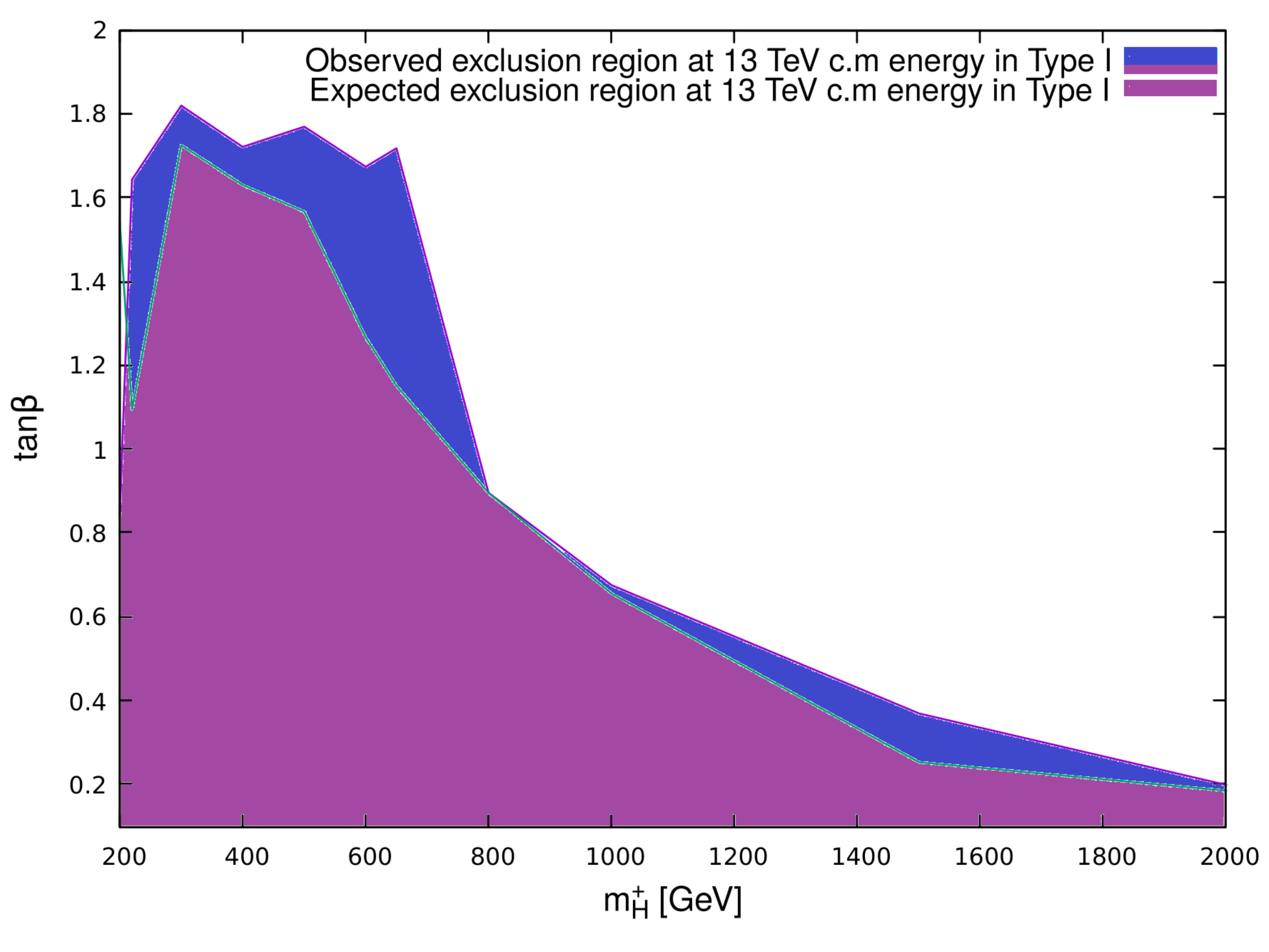}
   \caption{The comparison of expected (purple) and observed (blue plus purple) exclusion regions of charged Higgs parameter space in type1 2HDM at 13 TeV energy.}
    \label{fig10}
  \end{minipage}
\qquad
  \begin{minipage}[b]{0.4\textwidth}
    \includegraphics[width=8cm]{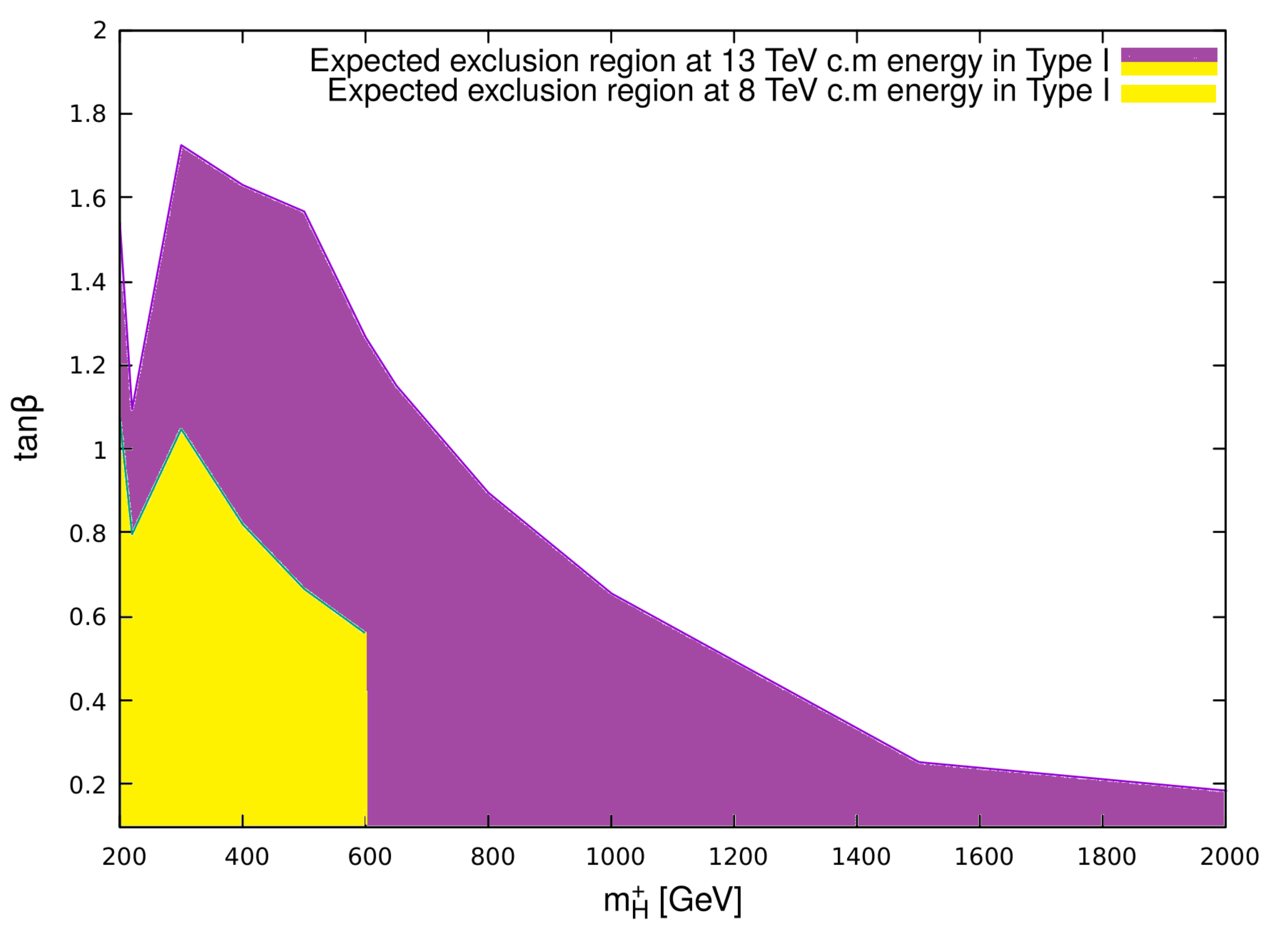}
\caption{The comparison of expected exclusion regions of charged Higgs parameter space in  type 1 2HDM at 8 TeV (yellow) and 13 TeV (purple plus yellow) energies}
    \label{fig11}
  \end{minipage}
  \qquad
  \begin{minipage}[b]{0.4\textwidth}
    \includegraphics[width=8cm]{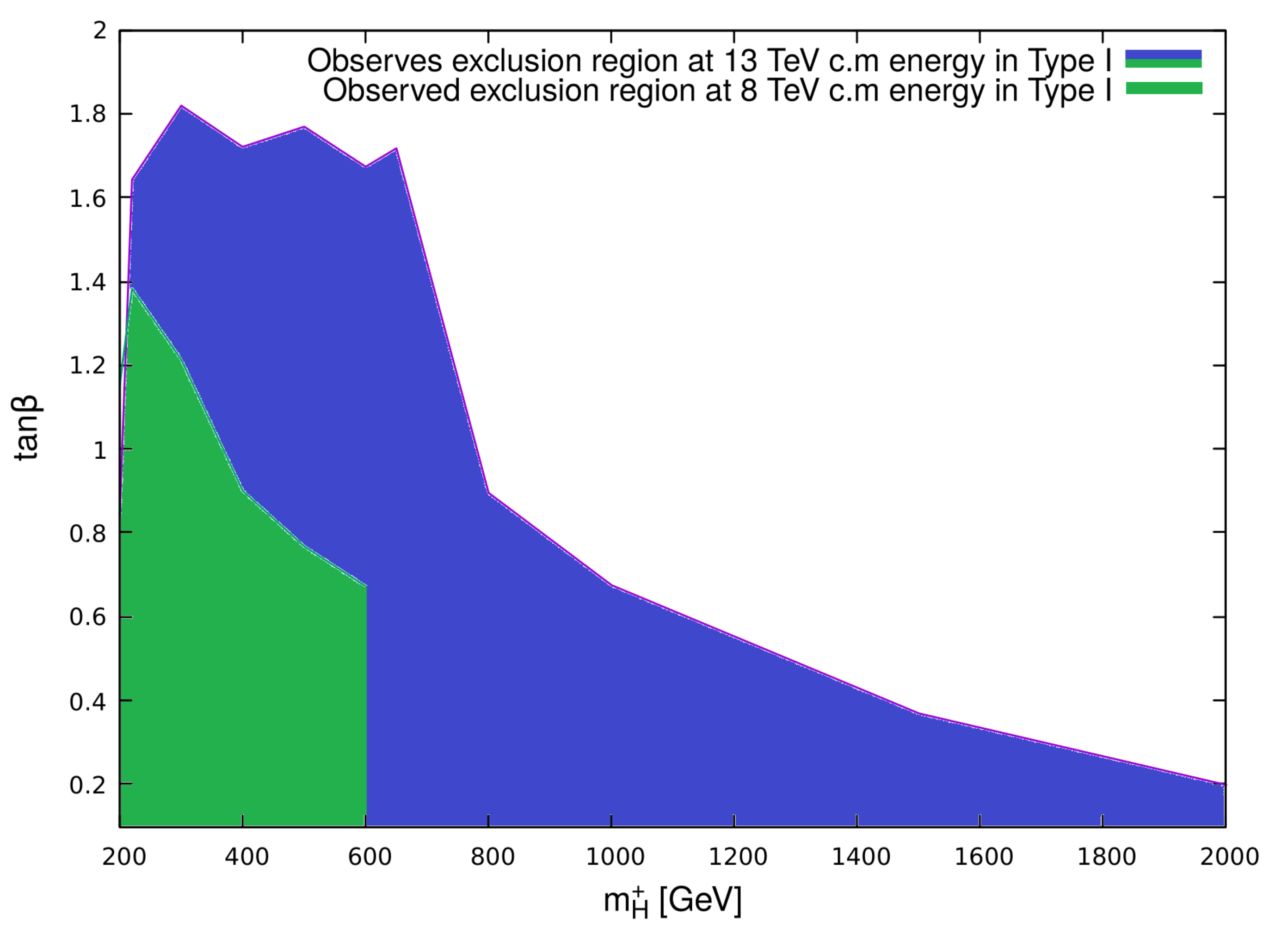}
\caption{The comparison of observed exclusion regions of charged Higgs parameter space in  type 1 2HDM at 8 TeV (green) and 13 TeV (blue plus green)  energies.}
    \label{fig12}
  \end{minipage}
    \qquad
  \begin{minipage}[b]{0.4\textwidth}
    \includegraphics[width=8cm]{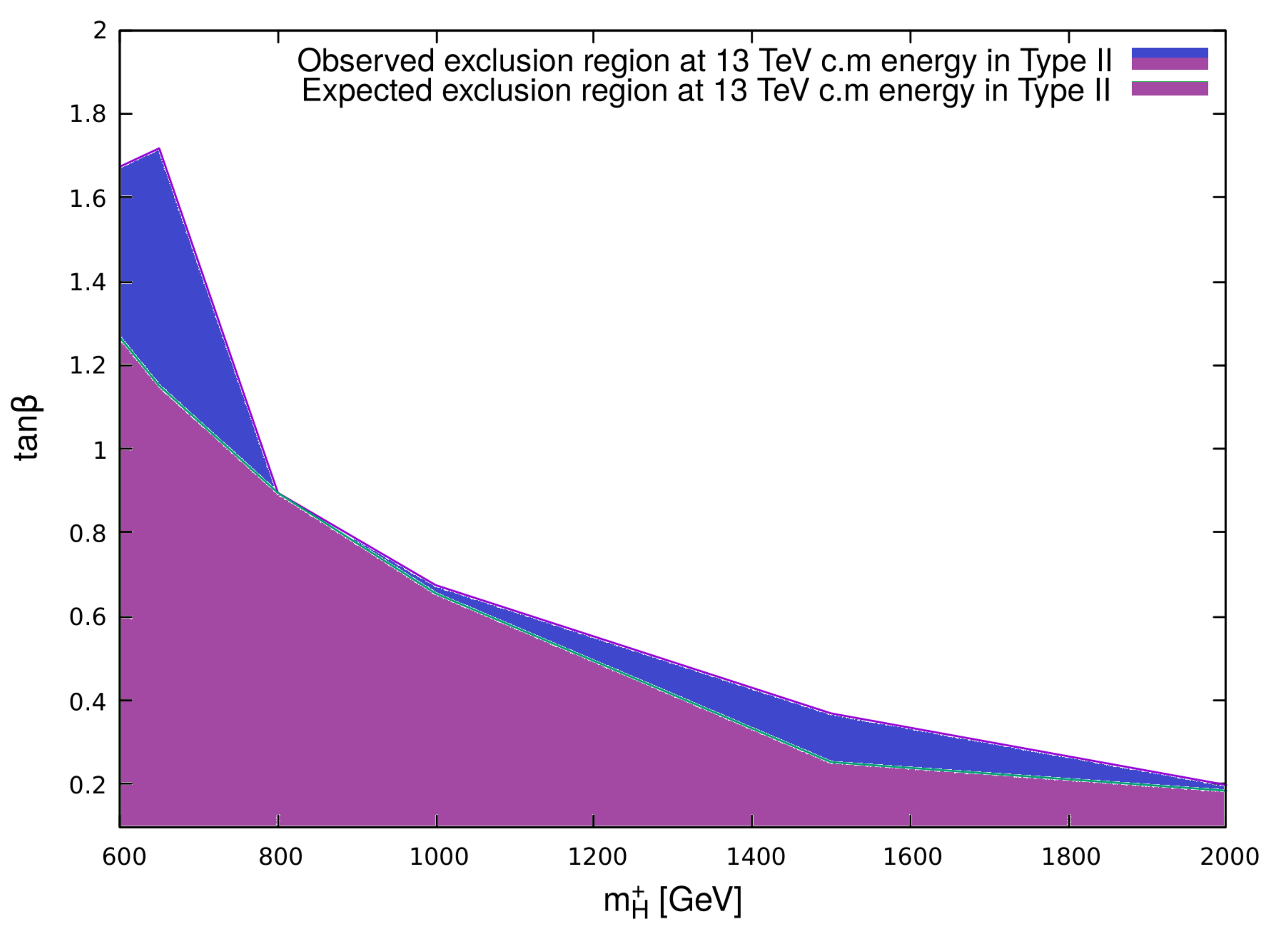}
   \caption{The comparison of expected (purple) and observed (blue plus purple) exclusion regions of charged Higgs parameter space in type II 2HDM at 13 TeV energy.}
    \label{fig13}
  \end{minipage}
\end{figure}

Here we put constraints on $\textcolor{red}{m_{H^\pm}} - \tan{\beta}$ space in 2HDM by using latest CMS results. These constraints on 2HDM parameter space are very significant. Computationally, $\sigma_{H^\pm} \times BR(H^\pm)$ has a non zero value for a vast 2HDM parameter space ie, $\sigma_{H^\pm} \times BR(H^\pm)$ has a non zero value for different values of $\textcolor{red}{m_{H^\pm}}$ and $\tan{\beta}$. which means that we can detect charged Higgs in wide 2HDM parameter space ie, we can discover charged Higgs in a vast 2HDM parameter space which is obviously a tough job. That is why it is necessary to put constraints or upper limits on such a wide 2HDM parameter space. These upper limits or constraints squeeze the 2HDM parameter space to a comparatively smaller region and then we can search for charged Higgs in a comparatively very smaller region of 2HDM parameter space which will reduce our labor work in discovering charged Higgs. Now the question is, how can we put such constraints or upper limits on 2HDM parameter space? The answer is CMS (Compact Muon Solenoid) experiment, \textcolor{red}{this experiment puts} upper limits on $\sigma_{H^\pm} \times BR(H^\pm)$ for various $\textcolor{red}{m_{H^\pm}}$ values.
The cross sections are available for various $\textcolor{red}{m_{H^\pm}}$ and $tan\beta$ values in \cite{b15}. The charged Higgs branching fractions are calculated by using 2HDMC for various $\textcolor{red}{m_{H^\pm}}$ and $\tan{\beta}$ values. It must be kept in mind that we are working in heavy Higgs scenario where the dominant decay channel is $H^+ \to t \bar{b}$  \textcolor{red}{for $\tan{\beta} < 10$} and is nearly equal to unity (one can easily see this from section 3). So we only take $H^+ \to t\bar{b}$ decay mode and  take it equal to 1  \textcolor{red}{for $\tan{\beta} < 10$}   and ignore the other fermionic decay modes because they are highly suppressed by this decay mode. Now in the product $\sigma_{H^\pm} \times BR(H^\pm \rightarrow tb^\pm)$ we are only left with $\sigma_{H^\pm} $ which means the sum of cross sections of positively and negatively charged Higgs ie, $\sigma_{H^\pm} =\sigma_{H^+} +\sigma_{H^-}$ . Now the cross sections in \textcolor{red}{\cite{b15}} are only for positive charged Higgs but we need $\sigma_{H^\pm}$, to overcome this problem multiply $\sigma_{H^+}$ with 2, as both types of charged Higgs have the same production cross sections. In short, $\sigma_{H^\pm} \times BR(H^\pm \rightarrow tb^\pm) =  2\sigma_{H^+}$ . Or just divide CMS $\sigma_{H^\pm}$ by 2 and then compare with the computational values of $\sigma_{H^+}$ we easily get $tan\beta$ value for each $\textcolor{red}{m_{H^\pm}}$.   
In CMS papers expected (median expected) upper limits and observed upper limits on $\sigma_{H^\pm} \times BR(H^\pm \rightarrow tb^\pm)$ are given for various $\textcolor{red}{m_{H^\pm}}$ values, both for 8 TeV \cite{b16} and 13 TeV \cite{b17} energies.\\
Now comparing the CMS results with our own computational results and using the interpolation formula, given in equation 5, we obtain the relationship between  $\textcolor{red}{m_{H^\pm}}$ and $tan\beta$ both for 8 TeV and 13 TeV and are given in tables number 11 and 12 respectively. Here we have two types of $\tan{\beta}$, "expected $\tan{\beta}$" which is obtained through the comparison of expected CMS limits with the computational values and "observed $\tan{\beta}$" which is obtained by the comparison of observed CMS limits with the computational values. Using the values of tables \ref{tab11} and \ref{tab12} we draw graphs between $\textcolor{red}{m_{H^\pm}}$ and $\tan{\beta}$ which are given in figures from \ref{fig9} to \ref{fig13}. 
	\begin{equation}
	\label{eq5}
\textcolor{red}{X - X_1 = \frac{(X_2 - X_1)(Y - Y_1)}{(Y_2 - Y_1)}}
	\end{equation}
	\textcolor{red}{In Eq. \ref{eq5} $(X_1,Y_1)$ and $(X_2,Y_2)$ represent two simulated data points \cite{b15} while $(X,Y)$ is the CMS data point laying in between $(X_1,Y_1)$ and $(X_2,Y_2)$. Here X is known but Y is unknown which can be found by using Eq. \ref{eq5}. In all data points, the abscissas denote $\sigma_{H^\pm}$ and ordinates denote $\tan{\beta}$. Repeating the same procedure, every time for a different mass of $H^{\pm}$ we obtain the values given in tables \ref{tab11} and \ref{tab12}}.
		
\end{multicols}

\begin{minipage}[b]{.40\textwidth}
 \centering
  \begin{tabular}{|l|l|l|} \hline
  $\textcolor{red}{m_{H^+}}$[GeV] & Expected $\tan{\beta}$ & Observed $\tan{\beta}$\\ \hline
  200 & 1.0743 & 1.1520\\ \hline
  220 & 0.7962 & 1.3828\\ \hline
  300 & 1.0472 & 1.2134\\ \hline
  400 & 0.8207 & 0.9006\\ \hline
  500 & 0.6661 & 0.7677\\ \hline
  600 & 0.5593 & 0.5781\\ \hline
  \end{tabular}
   \captionof{table}{$\tan{\beta}$ values for different values of $\textcolor{red}{m_{H^+}}$ at 8 TeV}
  \label{tab11}
\end{minipage}
\qquad
\begin{minipage}[b]{.40\textwidth}
 \centering
  \begin{tabular}{|l|l|l|} \hline
  $\textcolor{red}{m_{H^+}}$[GeV] & Expected $\tan{\beta}$ & Observed $\tan{\beta}$\\ \hline
  200 & 1.5414 & 0.8413\\ \hline
  220 & 1.0944 & 1.6426\\ \hline
  300 & 1.7250 & 1.8188\\ \hline
  400 & 1.6299 & 1.7206\\ \hline
  500 & 1.5663 & 1.7686\\ \hline
  600 & 1.2662 & 1.6735\\ \hline
  650 & 1.1509 & 1.7173\\ \hline
  800 & 0.8951 & 0.8952\\ \hline
  1000 & 0.6550 & 0.6749\\ \hline
  1500 & 0.2526 & 0.3691\\ \hline
  2000 & 0.1844 & 0.1984\\ \hline
  \end{tabular}
   \captionof{table}{$\tan{\beta}$ values for different values of $\textcolor{red}{m_{H^+}}$ at 13 TeV}
  \label{tab12}
\end{minipage}

%\begin{table}[h]
%\centering
%\begin{tabular}{|l|l|l|} \hline
%$m_H^+$[GeV] & Expected $\tan{\beta}$ & Observed $\tan{\beta}$\\ \hline
%200 & 1.0743 & 1.1520\\ \hline
%220 & 0.7962 & 1.3828\\ \hline
%300 & 1.0472 & 1.2134\\ \hline
%400 & 0.8207 & 0.9006\\ \hline
%500 & 0.6661 & 0.7677\\ \hline
%600 & 0.5593 & 0.5781\\ \hline
%\end{tabular}
%\caption{$\tan{\beta}$ values for different values of $m_H^+$ at 8 TeV}
%\label{tab11}
%\end{table}

%\begin{table}[h]
%\centering
%\begin{tabular}{|l|l|l|} \hline
%$m_H^+$[GeV] & Expected $\tan{\beta}$ & Observed $\tan{\beta}$\\ \hline
%200 & 1.5414 & 0.8413\\ \hline
%220 & 1.0944 & 1.6426\\ \hline
%300 & 1.7250 & 1.8188\\ \hline
%400 & 1.6299 & 1.7206\\ \hline
%500 & 1.5663 & 1.7686\\ \hline
%600 & 1.2662 & 1.6735\\ \hline
%650 & 1.1509 & 1.7173\\ \hline
%800 & 0.8951 & 0.8952\\ \hline
%1000 & 0.6550 & 0.6749\\ \hline
%1500 & 0.2526 & 0.3691\\ \hline
%2000 & 0.1844 & 0.1984\\ \hline
%\end{tabular}
%\caption{$\tan{\beta}$ values for different values of $m_H^+$ at 13 TeV}
%\label{tab12}
%\end{table}

\begin{multicols}{2}
\subsection{Results Discussion}
Figure \ref{fig9} shows the comparison of observed and expected exclusion regions in type I 2HDM at $\sqrt{s}=$ 8 TeV . It is obvious from figure \ref{fig9} that the observed exclusion region is always greater than the expected exclusion region in type I 2HDM at $\sqrt{s}=$ 8 TeV.  In figure \ref{fig10} we compare the expected and observed exclusion regions in type I 2HDM at $\sqrt{s} =$ 13 TeV. It is clear from figure \ref{fig10} that observed exclusion region is almost every where greater than the expected exclusion region in type I 2HDM at $\sqrt{s} =$ 13 TeV, except for the small region of $\textcolor{red}{m_{H^																										\pm}} = 200$ GeV to $\textcolor{red}{m_{H^\pm}} =$ 220 GeV in which the expected exclusion region is greater than the observed exclusion region.
We also equate the expected exclusion regions with the expected regions and observed exclusion regions with the observed exclusion regions at different energies. Figure \ref{fig11} shows the comparison of expected exclusion regions in Type I 2HDM at $\sqrt{s} =$8 and $\sqrt{s}=$13 TeV. Figure \ref{fig12} shows the comparison of observed exclusion regions at 8 TeV and 13 TeV energies. From both of these figures it is obvious that observed as well as expected exclusion regions for 13 TeV energy are always greater than 8 TeV energy in type I 2HDM i.e., 13 TeV puts more severe observed as well as expected upper limits on charged Higgs parameter space in type I 2HDM than 8 TeV energy. 
As in type II 2HDM the mass of $H^\pm$ is always greater than 580 GeV \textcolor{red}{\cite{b18}}, so we start from $\textcolor{red}{m_{H^\pm}}=$ 600 GeV we also see that the CMS upper limits end up on $\textcolor{red}{m_{H^\pm}} =$ 600 GeV for $\sqrt{s}=$8 TeV so, in type II 2HDM the CMS upper limits at $\sqrt{s}=$8 TeV have no involvement.
 The comparison of expected and observed exclusion regions in type II 2HDM at $\sqrt{s} =$ 13 TeV is given in figure \ref{fig13}. It is clear from figure 13, that the expected exclusion is always smaller than the observed exclusion region in type II 2HDM at 13 TeV energy.

Low $\tan{\beta}$ values have been discussed in our exclusion plots, for instance, take Fig.9 where $\tan{\beta}$ varies from 0.1 to 2. In CMS paper \cite{b16} the exclusion regions, for instance, take right column plots of Fig.11, have been plotted for large values of $\tan{\beta}$ and hence include some other exclusion regions too after $\tan{\beta} = 2$. That is why CMS exclusion regions are looking different from our exclusion regions. By comparing the CMS exclusion plot of the figure under consideration with our exclusion plot of Fig.9, it can be found that the peaks of exclusion regions in both cases reach to $\tan{\beta} = 1.4$. It means that if CMS used lower values of $\tan{\beta}$ as we did, the CMS exclusion regions would perfectly match our exclusion regions.    
We used lower values of $\tan{\beta}$ because during the comparison of simulated values with the CMS values all $\tan{\beta}$ values turned out to be lower than 2 through interpolation formula.
\section{Conclusion}

CMS upper limits on $H^\pm$ parameter space in 2HDM are very significant, because we have a vast 2HDM parameter space and computationally $\sigma_{H^\pm} \times BR(H^\pm)$ has a non zero value through out this vast parameter space ie, the discovery of charged Higgs is possible in a vast 2HDM parameter space which is really a headache. To overcome this problem we need to restrict the 2HDM parameter space so that the observability chances of charged Higgs could be enhanced in a comparatively small region which is obviously an easier work than searching for charged Higgs in 2HDM parameter space which is not constrained. The reduction of the size of 2HDM parameter space ($H^\pm$ parameter space in 2HDM) is provided through the CMS experiment upper limits. We discussed two types of CMS upper limits, one expected upper limits and other observed upper limits at $\sqrt{s}=$8 TeV and at $\sqrt{s}=$13 TeV. We concluded that the observed exclusion region is always greater than expected exclusion region in type I and type II 2HDM at $\sqrt{s}=$8 TeV and $\sqrt{s}=$13 TeV, except at $\textcolor{red}{m_{H^\pm}}$ = 200 GeV to $\textcolor{red}{m_{H^\pm}}$ = 220 GeV only at $\sqrt{s}=$13 TeV, where the expected exclusion region of charged Higgs parameter space is slightly greater than the observed one.
We also observe and conclude that $\sqrt{s}=$13 TeV excludes more charged Higgs parameter space (expected as well as observed exclusion regions) in all types of 2HDM than $\sqrt{s}=$8 TeV.\\
 
\end{multicols} 

\begin{multicols}{2}

\end{multicols} 
\end{CJK*}
\end{document}